\documentclass{ws-procs9x6}

\begin{document}

\title{Chiral quark model spin filtering mechanism and hyperon polarization}

\author{S.~M. TROSHIN and N.~E. TYURIN}

\address{Institute
for High Energy Physics,\\
Protvino, Moscow Region, 142281 Russia}

\maketitle

\abstracts{The model combined with unitarity and impact parameter picture  provides
simple mechanism for generation of hyperon polarization  in collision of unpolarized hadrons.
We  concentrate  on a particular problem of $\Lambda$-hyperon polarization and derive
its linear $x_F$-dependence as well as its energy and transverse momentum
independence at large $p_\perp$ values. Mechanism
 responsible for the single--spin asymmetries  in pion production is also discussed.
}

One of the most interesting and persistent for a long time spin phenomena
was observed in inclusive hyperon production in collisions of
unpolarized hadron beams. A very significant polarization of
$\Lambda$--hyperons has been discovered almost three  decades ago\cite{newrev}.
  Experimentally the process
of $\Lambda$-production has been studied more extensively than other hyperon
production processes.
Therefore we will emphasize on the particular riddle  of $\Lambda$--polarization
because  spin structure of this particle is most simple and
 is determined by strange quark only. This mechanism can also
be used for the explanation of single-spin asymmetries in the inclusive pion production.

It should be noted that understanding of
transverse single-spin asymmetries in DIS (in contrast to the hyperon polarization)
has observed significant progress during last years; this progress is related
to an account of final-state interactions from gluon exchange\cite{brodsky,metz} --
coherent effect not suppressed in the Bjorken limit.

Experimental
 situation with hyperon polarization is widely known and stable for a long time.
Polarization of $\Lambda$ produced in the unpolarized inclusive $pp$--interactions
is negative and energy
independent. It increases linearly with $x_F$ at large transverse momenta
($p_\perp\geq 1$ GeV/c),
and for such
values of transverse momenta   is almost
$p_\perp$-independent\cite{newrev}.

On the theoretical side,  perturbative QCD
with a straightforward collinear factorization scheme
leads to small values of $\Lambda$--polarization
\cite{pump,gold} which are far below of the corresponding experimental data.
Modifications of this scheme and  account for higher twists contributions allows
to obtain higher magnitudes of polarization but do not change
a decreasing  dependence proportional to
$p_\perp^{-1}$ at large transverse momenta\cite{efrem,sterm,koike}.
It is difficult to reconcile this behavior  with the flat  experimental
data dependence on the transverse momenta. Inclusion of the internal transverse momentum
of partons ($k_\perp$--effects) into the
polarizing fragmentation functions  leads   to similarly decreasing polarization\cite{anselm}.
In addition it should be noted that the perturbative QCD has also problems in the description
of the unpolarized scattering, e.g. in inclusive cross-section
for $\pi^0$-production, at the energies lower than the RHIC energies\cite{bsof}.

The essential point  of the approaches mentioned above is that the vacuum at short distances
is taken to be a perturbative one.
There is an another possibility. It might happen  that
the polarization dynamics in strangeness production originates from the genuine
nonperturbative sector of QCD (cf. e.g.\cite{spin02}).
In the nonperturbative sector of QCD the  two important
phenomena,  confinement and spontaneous breaking of chiral symmetry ($\chi$SB)\cite{mnh}
should be reproduced.
The  relevant scales   are characterized by the
parameters $\Lambda _{QCD}$ and $\Lambda _\chi $, respectively.  Chiral $SU(3)_L\times
SU(3)_R$ symmetry is spontaneously  broken  at the distances
in  the range between
these two scales.  The $\chi$SB mechanism leads
to generation of quark masses and appearance of quark condensates. It describes
transition of current into  constituent quarks.
  Constituent quarks are the quasiparticles, i.e. they
are a coherent superposition of bare  quarks, their masses
have a magnitude comparable to  a hadron mass scale.  Therefore
hadron  is often represented as a loosely bounded system of the
constituent quarks.
These observations on the hadron structure lead
to  understanding of several regularities observed in hadron
interactions at large distances. It is well known  that such picture  provides
reasonable  values  for the static characteristics of hadrons, for
 instance, their magnetic moments. The other well known direct result
   is  appearance of the Goldstone bosons.

The most recent  approach  to
single--spin asymmetries (SSA) based on nonperturbative QCD has been developed in\cite{burk} where, in particular,
$\Lambda$-polarization has been related to the large magnitude of the transverse
 flavor dipole moment of the transversely polarized baryons in the infinite momentum frame.
  It is based
on the parton picture in the impact parameter space and assumed specific helicity--flip
generalized parton distribution.

The instanton--induced mechanism of SSA generation was considered in\cite{koch,shur}
and relates those asymmetries with a genuine nonperturbative QCD interaction.
It should be noted that the physics of instantons (cf. e.g.\cite{inst})
can provide microscopic explanation for the $\chi$SB mechanism.

We discuss here mechanism for hyperon polarization based on chiral
quark model\footnote{It  has been successfully applied
for the  explanation of the nucleon spin structure\cite{cheng}.}\cite{mnh} and the
filtering spin states related to unitarity in the $s$-channel.
This mechanism connects polarization with  asymmetry in the
position (impact parameter) space.

As it was already mentioned constituent quarks and Goldstone bosons are the effective
degrees of freedom in the chiral quark model. We consider a
 hadron consisting of the valence
constituent quarks located in the central core which is embedded into  a quark
condensate. Collective excitations of the condensate are the Goldstone bosons
and the constituent quarks interact via exchange
of Goldstone bosons; this interaction is mainly due to a pion field which is of the flavor--
 and spin--exchange nature. Thus, quarks generate a strong field which
binds them\cite{diak}.

At the first stage of hadron interaction common effective
self-consistent field is appeared.
Valence constituent quarks   are
 scattered simultaneously (due to strong coupling with Goldstone bosons)
and in a quasi-independent way by this effective strong
 field. Such ideas were already used in the model\cite{csn} which has
been applied to description of elastic scattering and hadron production\cite{mult}.

The initial state  particles (protons) are unpolarized.
It means that states with spin up and spin down have equal probabilities.
The main idea of the proposed mechanism is the  filtering
of the two initial spin states of equal probability due to different strength of interactions. The particular
mechanism of such filtering can be developed on the basis of chiral quark model,
formulas for inclusive cross section (with account for the unitarity)\cite{tmf} and
notion on the quasi-independent nature of valence quark scattering in the effective field.

We will exploit the feature of chiral quark model that constituent quark $Q_\uparrow$
with transverse spin in up-direction can fluctuate into Goldstone boson and
  another constituent quark $Q'_\downarrow$ with opposite spin direction,
   i. e. perform a spin-flip transition\cite{cheng}:
\begin{equation}\label{trans}
Q_\uparrow\to GB+Q'_\downarrow\to Q+\bar Q'+Q'_\downarrow.
\end{equation}
An absence of arrows means that the corresponding quark is unpolarized.
To compensate quark spin flip $\delta {\bf S}$ an orbital angular momentum
$\delta {\bf L}=-\delta {\bf S}$ should be generated in final state of reaction (\ref{trans}).
The presence of this orbital momentum $\delta {\bf L}$  in its turn
means  shift in the impact parameter
value of the final quark $Q'_\downarrow$ (which is transmitted to the shift in the impact
parameter of $\Lambda$)
\[
\delta {\bf S}\Rightarrow\delta {\bf L}\Rightarrow\delta\tilde{\bf b}.
\]
Due to   different strengths of interaction at the different values of the
impact parameter, the processes of transition to the
spin up and down states will have different probabilities which  leads eventually to
polarization of $\Lambda$.

\begin{figure}[h]
\begin{center}
  \resizebox{4cm}{!}{\includegraphics*{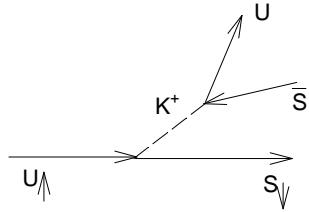}}
\end{center}
\caption{Transition of the spin-up constituent quark $U$ to the spin-down strange quark.
 \label{ts1}}
\end{figure}
In a particular case of $\Lambda$--polarization the relevant transitions
of constituent quark $U$ (cf. Fig. 1) will be correlated with the shifts $\delta\tilde b$
in impact parameter $\tilde b$ of the final
$\Lambda$-hyperon, i.e.:
\begin{eqnarray}
  \nonumber U_\uparrow & \to & K^+ + S_\downarrow\Rightarrow\;\;-\delta\tilde{\bf b} \\
\label{spinflip} U_\downarrow & \to & K^+ + S_\uparrow\Rightarrow\;\;+\delta\tilde {\bf b}.
\end{eqnarray}
Eqs. (\ref{spinflip}) clarify mechanism of the filtering of spin states:
 when shift in impact
parameter is $-\delta\tilde {\bf b}$ the
interaction is stronger compared to the case when shift is $+\delta\tilde {\bf b}$,
and the final $S$-quark
(and $\Lambda$-hyperon) is polarized negatively. Thus, the particular mechanism of
 filtering of spin states is related to the
 emission of Goldstone bosons by constituent quarks.

It is important to note here that the shift of $\tilde{\bf b}$
(the impact parameter of final hyperon)
is translated to the shift of the impact parameter of the initial particles according
to the relation between impact parameters in the multiparticle production\cite{webb}:
\begin{equation}\label{bi}
{\bf b}=\sum_i x_i{ \tilde{\bf  b}_i}.
\end{equation}
The variable $\tilde b$ is conjugated to the transverse momentum of $\Lambda$,
but relations  between functions depending on the impact parameters
$\tilde b_i$ are nonlinear.
We consider production of $\Lambda$ in the fragmentation region, i.e.
at large $x_F$ and therefore use approximate relation
\begin{equation}\label{bx}
b\simeq x_F\tilde b,
\end{equation}
which results from Eq. (\ref{bi})\footnote{We make here an additional assumption on the
small values of Feynman $x$ for other particles}.

The explicit formulas for inclusive
cross--sections of the process
\[ h_1 +h_2\rightarrow h_3^\uparrow +X, \] where hadron $h_3$ is a hyperon whose
transverse polarization is measured were obtained in
\cite{tmf}. The main feature of this formalism is an account of
unitarity in the direct channel of
reaction. The corresponding formulas have the form
\begin{equation}
{d\sigma^{\uparrow,\downarrow}}/{d\xi}= 8\pi\int_0^\infty
bdb{I^{\uparrow,\downarrow}(s,b,\xi)}/ {|1-iU(s,b)|^2},\label{un}
\end{equation}
where $b$ is the impact  parameter of the initial
particles. Here the function
$U(s,b)$ is the generalized reaction matrix (for unpolarized scattering)
which is determined by the basic dynamics of elastic scattering.

The functions $I^{\uparrow,\downarrow}$ in Eq. (\ref{un}) are related   to the
functions  $|U_n|^2$, where $U_n$  are the multiparticle
analogs of the $U$(cf. \cite{tmf}). The kinematical variables $\xi$
($x_F$ and $p_\perp$) describe the state of the produced particle
$h_3$.
   Arrows $\uparrow$ and $\downarrow$ denote
   transverse spin directions of the final hyperon $h_3$.

Polarization
can be expressed in terms of the functions $I_{-}$, $I_{0}$ and $U$:
\begin{equation} P(s,\xi)=\frac{\int_0^\infty bdb
I_-(s,b,\xi)/|1-iU(s,b)|^2} {2\int_0^\infty bdb
I_0(s,b,\xi)/|1-iU(s,b)|^2},\label{xnn}
\end{equation}
where $I_0=1/2(I^\uparrow+I^\downarrow)$ and $I_-=(I^\uparrow-I^\downarrow)$.

On the basis of the described chiral quark filtering mechanism we can
assume that the functions
$I^\uparrow(s,b,\xi)$ and $I^\downarrow(s,b,\xi)$ are related to the functions
$I_0(s,b,\xi)|_{\tilde b+\delta\tilde b }$ and $I_0(s,b,\xi)|_{\tilde b-
\delta\tilde {b} }$,
respectively, i.e.
\begin{equation}\label{der}
I_-(s,b,\xi)=I_0(s,b,\xi)|_{\tilde {b}+\delta\tilde {b} }-
I_0(s,b,\xi)|_{\tilde{b}-\delta\tilde{b} }
=2\frac{\delta I_0(s,b,\xi)}{\delta\tilde{b}}\delta\tilde b.
\end{equation}

We can connect $\delta\tilde b$ with the radius of quark interaction
$r_{U\to S}^{flip}$
responsible for the transition $U_\uparrow\to S_\downarrow$ changing quark spin and flavor:
\[
\delta\tilde b\simeq r_{U\to S}^{flip}.
\]

The following expression for polarization $P_\Lambda(s,\xi)$ can be obtained
\begin{equation} P_\Lambda(s,\xi)\simeq x_Fr_{U\to S}^{flip}\frac{\int_0^\infty bdb
I'_0(s,b,\xi)db/|1-iU(s,b)|^2} {\int_0^\infty bdb
I_0(s,b,\xi)/|1-iU(s,b)|^2},\label{poll}
\end{equation}
where $I'_0(s,b,\xi)={dI_0(s,b,\xi)}/{db}$.

It is clear that
polarization of $\Lambda$ - hyperon (\ref{poll})
should be negative because $I'_0(s,b,\xi)<0$.

The generalized
reaction matrix $U(s,b)$ (in a pure imaginary case) is
the following
\begin{equation} U(s,b) = i\tilde U(s,b)=ig(s)\exp(-Mb/\zeta ),
\,\,
 g(s)\equiv  g_0\left [1+\alpha
\frac{\sqrt{s}}{m_Q}\right]^N,
 \label{x}
\end{equation}
$M$ is the total mass of $N$ constituent quarks with mass $m_Q$ in
the initial hadrons; $\alpha$ and $g_0$ are the parameters of
model. Parameter $\zeta$ is the one which determines a universal scale for
the quark interaction radius, i.e. $r_Q=\zeta /m_Q$.
To evaluate polarization dependence on $x_F$ and $p_\perp$
we use semiclassical correspondence  between transverse momentum and impact parameter
  values.Choosing the region of  small $p_\perp$ we select the large values of impact parameter
 and therefore  we  have
\begin{equation} P_\Lambda(s,\xi)\propto -x_Fr_{U\to S}^{flip}
\frac{M}{\zeta}\frac{\int_{b>R(s)} bdb
I_0(s,b,\xi)\tilde U(s,b) } {\int_{b>R(s)} bdb
I_0(s,b,\xi)},\label{pollsm}
\end{equation}
where $R(s)\propto \ln s$ is the hadron interaction radius, which serve as a scale
of large and small impact parameter values.
At large values of impact parameter $b$:
$\tilde U(s,b)\ll 1$ for $b\gg R(s)$ and therefore
 we will have small polarization $P_\Lambda\simeq 0$
in the region of small and moderate $p_\perp\leq 1$ GeV/c.
At small values of $b$ (and large $p_\perp$): $\tilde U(s,b)\gg 1$
 and the following approximate relations
are valid
\begin{equation}
\int_{b<R(s)} bdb\frac{I_0(s,b,\xi)\tilde U(s,b)}{ [1+\tilde U(s,b)]^{3}}
\simeq
\int_{b<R(s)} bdb
I_0(s,b,\xi)\tilde U(s,b)^{-2}\label{simil},
\end{equation}
since we can neglect unity in the denominators of the integrands.
\begin{figure}[htb]
\begin{center}
  \resizebox{4cm}{!}{\includegraphics*{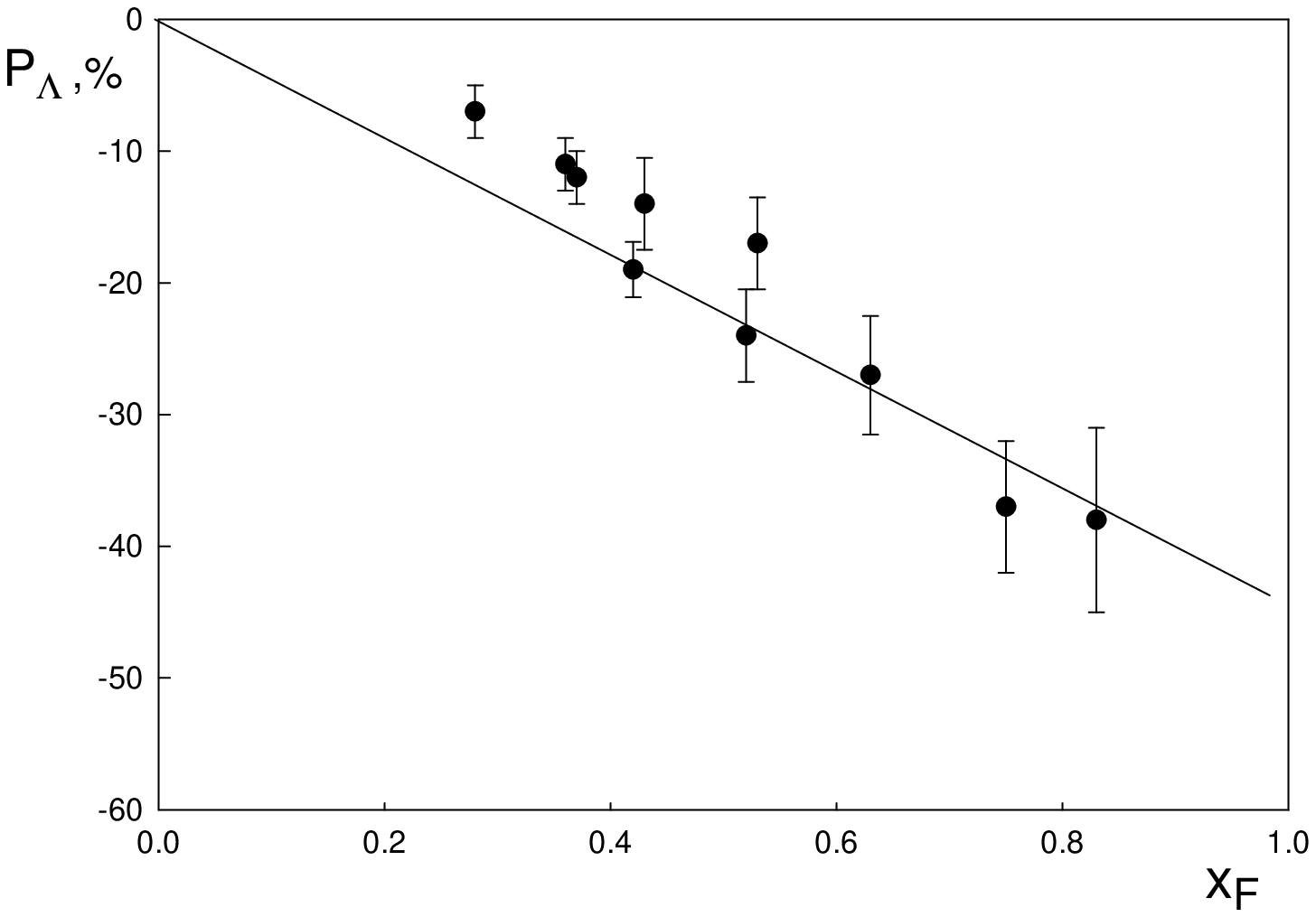}}\;\;\quad
  \resizebox{4cm}{!}{\includegraphics*{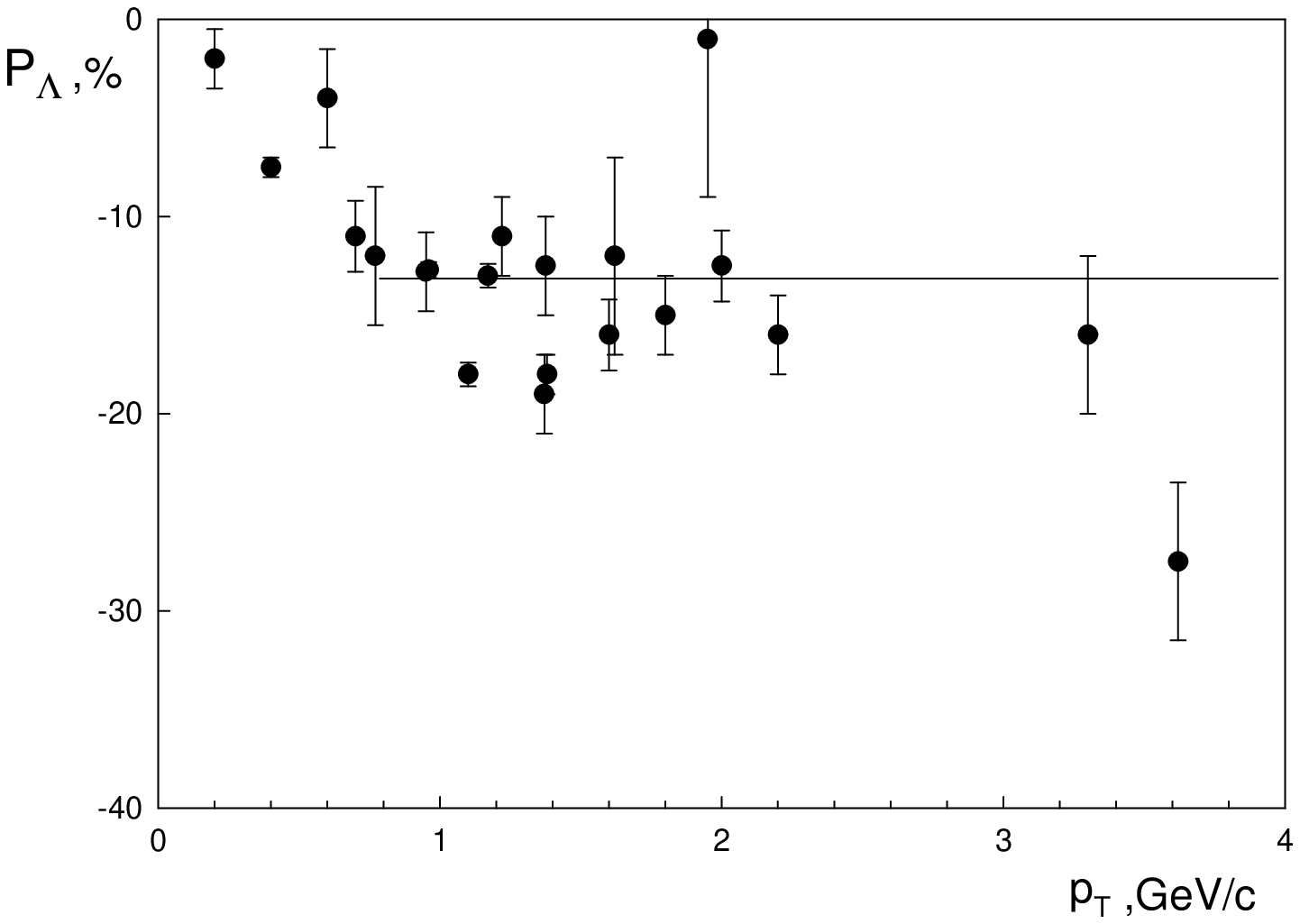}}
\end{center}
\caption{$x_F$ (left panel) and $p_T$ (right panel)
 dependencies of the $\Lambda$-hyperon
polarization} \label{ts}
\end{figure}
Thus,  the energy and $p_\perp$-independent behavior
of polarization $P_\Lambda$ takes place at large values of $p_\perp$:
\begin{equation} P_\Lambda(s,\xi)\propto -x_Fr_{U\to S}^{flip}
{M}/\zeta.\label{polllg}
\end{equation}
This flat transverse momentum dependence results from the similar
rescattering effects for the different spin states.
The numeric value of polarization $P_\Lambda$ can be large: there are
no small factors in (\ref{polllg}). In (\ref{polllg}) $M$ is
proportional to two nucleon masses, the value of parameter $\zeta \simeq 2$.
 We expect that $r_{U\to S}^{flip}\simeq
0.1-0.2$ fm on the basis of the model\cite{csn,tmf}, however,
this is a crude estimate. The above qualitative
 features of polarization dependence on $x_F$,
$p_\perp$ and energy are in a good agreement with the experimentally observed trends\cite{newrev}.
For example, Fig. 2 demonstrates that the linear $x_F$ dependence is in a good agreement with
the experimental data in the fragmentation region ($x_F\geq 0.4$) where the model
should work. Of course,
the conclusion on the $p_\perp$--independence of polarization is a rather approximate one
and deviation from such behavior cannot be excluded.

The proposed mechanism deals with effective degrees of freedom and takes into
account collective aspects of QCD dynamics. Together with unitarity, which is an essential
ingredient of this approach, it allows  to
 obtained results for polarization dependence on kinematical variables
 in  agreement with the  experimental  behavior
of $\Lambda$-hyperon polarization, i.e.
 linear dependence on $x_F $ and
flat dependence
on $p_\perp$ at large $p_\perp$
in the fragmentation region are reproduced.
Those dependencies together with the energy independent
behavior of polarization at large transverse
momenta are the straightforward consequences of this model.
We discussed here  particle production in the fragmentation region.
In the central region where correlations
 between impact parameter of the initial and impact parameters of the final particles
 being weakened, the polarization cannot be generated due to chiral quark filtering
 mechanism.
Moreover, it is  clear that since antiquarks are produced through spin-zero Goldstone bosons
we should expect $P_{\bar\Lambda}\simeq 0$.
The chiral quark filtering is also relatively suppressed when compared to direct elastic
 scattering of quarks in effective   field and therefore
   should not play a role in the reaction $pp\to pX$ in the fragmentation
 region, i.e. protons should be produced unpolarized. These features take place
 in the experimental data set.
The
application
of this mechanism to description of polarization of other hyperons is more complicated
problem,
since they
could have two or three strange quarks and spins of  $U$ and $D$
quarks can also make contributions into their polarizations.
Finally, it was shown that the mechanism reversed to chiral
quark filtering can provide description of the SSA in $\pi^0$ production
measured at FNAL and recently at RHIC in the fragmentation region and it leads to the
energy independence of the  asymmetry.

One of the authors (S.T.) is very grateful to the Organizing Committee of
``Transversity 2005'' for the warm hospitality in Como during this very interesting workshop.

\end{document}